\documentclass[12pt]{article}  
\usepackage{arxiv}             
\usepackage{cite}
\usepackage{amsmath,amssymb,amsfonts}
\usepackage{algorithmic}
\usepackage{graphicx}
\usepackage{textcomp}
\usepackage{xcolor}
\usepackage{float} 

\usepackage{comment}
\usepackage[table]{xcolor}
\usepackage{multirow} 
\usepackage{booktabs}
\usepackage{url}

\usepackage{fancyhdr}
\def\BibTeX{{\rm B\kern-.05em{\sc i\kern-.025em b}\kern-.08em
    T\kern-.1667em\lower.7ex\hbox{E}\kern-.125emX}}
\begin{document}

\title{Soundscapes in Spectrograms: Pioneering Multilabel \\
Classification for South Asian Sounds
\thanks{Amygdala AI is an international volunteer-run research group that advocates for \textit{AI for a better tomorrow}, accessible at \protect\url{http://amygdalaai.org/}}
}

\author{
\large \textbf{Sudip Chakrabarty}\textsuperscript{1}, \large \textbf{ Pappu Bishwas}\textsuperscript{1}, \large \textbf{ Rajdeep Chatterjee}\textsuperscript{1},\\
\large \textbf{Tathagata Bandyopadhyay}\textsuperscript{2}, \large \textbf{ Digonto Biswas}\textsuperscript{1}, \large \textbf{ Bibek Howlader}\textsuperscript{3} \\
\\
\textsuperscript{1}School of Computer Engineering, KIIT Deemed to be University, Bhubaneswar, India \\
\textsuperscript{2}School of Computation, Information and Technology, Technical University of Munich, Germany \\
\textsuperscript{3}School of Computer Engineering, American International University, Dhaka, Bangladesh \\
\\
Emails: \{sudipchakrabarty6, cse.rajdeep, gata.tatha14\}@gmail.com
}

\maketitle

\begin{abstract}
    Environmental sound classification is a field of growing importance for urban monitoring and cultural soundscape analysis, especially within the acoustically rich environments of South Asia. These regions present a unique challenge as multiple natural, human, and cultural sounds often overlap, straining traditional methods that frequently rely on Mel Frequency Cepstral Coefficients (MFCC). This study introduces a novel spectrogram-based methodology with a superior ability to capture these complex auditory patterns. A Convolutional Neural Network (CNN) architecture is implemented to solve a demanding multilabel, multiclass classification problem on the SAS-KIIT dataset. To demonstrate robustness and comparability, the approach is also validated using the renowned UrbanSound8K dataset. The results confirm that the proposed spectrogram-based method significantly outperforms existing MFCC-based techniques, achieving higher classification accuracy across both datasets. This improvement lays the groundwork for more robust and accurate audio classification systems in real-world applications.
\end{abstract}


\section{\textbf{Introduction}}\label{sec1}
The analysis of environmental sounds is crucial for urban surveillance, public safety, and preserving cultural soundscapes. In South Asia, this task is even more challenging due to its rich and dynamic auditory environment, where natural, human-made, and cultural sounds often blend together. Accurately identifying and categorizing these sounds is essential for developing intelligent urban systems, improving environmental awareness, and preserving the region’s acoustic heritage. Moreover, environmental audio carries vital contextual information that can be used to monitor real-time changes, detect anomalies, and support decision-making in resource-limited settings. However, the complexity of South Asian soundscapes, with multiple overlapping sound sources, poses significant challenges for traditional classification methods.

Traditional methods for classifying mixed environmental sounds often rely on Blind Source Separation (BSS) techniques, such as ICA and PCA \cite{5749353,7877996}, which require prior knowledge of the number of sources. Another conditional audio separation approach \cite{bandyopadhyay2024spectron} could have been used, but such methods require clean, pre-recorded samples of each sound class. These techniques perform poorly in real-world, dynamic environments where overlapping signals are unknown or numerous. In addition, most existing solutions are trained on narrowly focused datasets, limiting their adaptability to diverse and unpredictable acoustic settings.

To overcome these challenges, a spectrogram-driven deep learning method is introduced for handling multi-label and multi-class classification of mixed audio samples. Unlike MFCC-based methods \cite{9579881}, which often struggle to capture fine-grained temporal and frequency variations, the proposed spectrogram-based CNN model is able to learn complex audio patterns directly, without requiring explicit source separation. This makes the model highly suitable for diverse and overlapping soundscapes commonly found in the South Asian region.

The method is evaluated on an expanded version of the SAS-KIIT dataset \cite{10829485}, which now includes 21 South Asian sound classes, it is also tested on the widely used UrbanSound8K dataset \cite{Salamon:UrbanSound:ACMMM:14} for comparison. The spectrogram-based method demonstrated noticeably better performance than traditional MFCC-based approaches, with higher classification accuracy observed across both datasets. This demonstrates how spectrogram analysis contributes to more accurate environmental sound recognition and supports the documentation of South Asia’s diverse acoustic environments.

This paper is structured as follows: Section \ref{sec3} details the dataset utilized and the preprocessing techniques applied. Section \ref{sec4} then introduces the proposed methodology. Section \ref{sec5} presents and discusses the experimental results, emphasizing the key findings. The paper concludes in Section \ref{sec6} with a summary of the work and suggestions for future research directions.

\vspace{2mm}
\section{\textbf{Literature Survey and Related Works}}\label{sec2}
Many papers have implemented MFCC and spectrogram-based methods for various audio classification tasks. In \cite{sharma20_interspeech}, an attention-based CNN model was introduced for environmental sound classification, utilizing multiple feature types such as MFCC, GFCC, CQT, and Chromagram. In \cite{8047110}, SVM with a Gaussian kernel and feature extraction techniques, including short-time energy, zero-crossing rate, and MFCC, was used for audio classification.
Several studies have employed convolutional neural networks (CNNs)\cite{10829485} in conjunction with spectrograms and Mel-Frequency Cepstral Coefficients (MFCC) for sound classification, demonstrating significant improvements in model performance across various audio recognition tasks. 
In the field of multi-label classification \cite{article1}, many techniques have been explored and evaluated across various datasets to assess their performance and efficiency. Tasnim Akter et al.\cite{10685563} propose a multi-label sound classification approach using deep learning models, including CNN, LSTM \cite{inproceedings}, and GRU, for instrument recognition from audio signals. This paper\cite{a24_interspeech} presents multi-label classification uisng Mel-spectrograms and a deep CNN called Mel-Graph-GCN \cite{Wang21} framework for bird species classification from field recordings. Despite significant progress in environmental sound classification, most studies rely on culturally limited datasets and focus primarily on single-label tasks, overlooking the complexities of multilabel scenarios. In response to these limitations, our study focuses on developing a spectrogram-based CNN framework tailored for multilabel classification within the rich and chaotic soundscapes of South Asia, using both regional and global datasets to ensure broader applicability and robustness.

\vspace{2mm}
\section{\textbf{Datasets and Preprocessing}}\label{sec3}
In this study, two datasets are used to evaluate our approach. SAS-KIIT\footnote{SAS-KIIT: https://sas-kiit.netlify.app/} captures diverse South Asian sounds, including natural, urban, and human-made events. UrbanSound8K\footnote{UrbanSound8K: https://urbansounddataset.weebly.com/urbansound8k.html}, a benchmark dataset, features various urban noise types, making it ideal for assessing classification accuracy in complex environments. The dataset details and preprocessing steps are outlined below.

\subsection{\textbf{SAS-KIIT Dataset}}
The SAS-KIIT dataset is a curated collection of South Asian soundscapes, featuring auditory events from countries such as India, Bangladesh, Nepal, Bhutan and Afghanistan for multiclass sound classification tasks.

\vspace{1mm}
\subsubsection{Dataset Overview}
The dataset is a comprehensive collection of auditory events that were obtained from different environments, such as urban, rural, and industrial settings, offering a broad representation of real-world acoustic conditions. To make the dataset even more complete, some rare environmental sounds were added, improving its diversity. This extended version of the dataset includes the following sound classes:

\begin{table}[h]
    \centering
    \caption{List of 21 Sound Classes in the SAS-KIIT Dataset}
    \begin{tabular}{ll}
        \toprule 
        1. Tanpura & 12. Launch Engine \\
        2. Traditional Song & 13. Flute \\
        3. Railway Engine & 14. Buddhist Prayer \\
        4. Children Class Noise & 15. Fish Market \\
        5. Harmonium & 16. Tiger \\
        6. Dhak & 17. Elephant \\
        7. Tabla & 18. Kalboishakhi Storm \\
        8. Azan & 19. Sanatan Religion Aroti \\
        9. Church Prayer & 20. Rickshaw Horn \\
        10. Irrigation Engine & 21. Afghanistan Pashto Music \\
        11. Ektara & \\
        \bottomrule 
    \end{tabular}
    \label{tab:sound_classes}
\end{table}

\vspace{1mm}
\subsubsection{Organization and Characteristics}
The dataset is systematically organized into 10 folders, ensuring a balanced distribution of audio samples for training and evaluation. Each sound class contains 450 segments, each lasting 4 seconds, resulting in a total of 9,450 audio segments. The audio files are in WAV format with a sampling rate of 44.1 kHz. Along with the audio files, metadata annotations are included, detailing essential information such as segment names, slicing start and end times, class IDs, class names, and folder IDs. This structured organization facilitates efficient data handling and model training.

\begin{figure}[h]
    \centering
    \begin{minipage}[]{0.49\textwidth} 
        \centering
        \includegraphics[width=\textwidth, height=6cm]{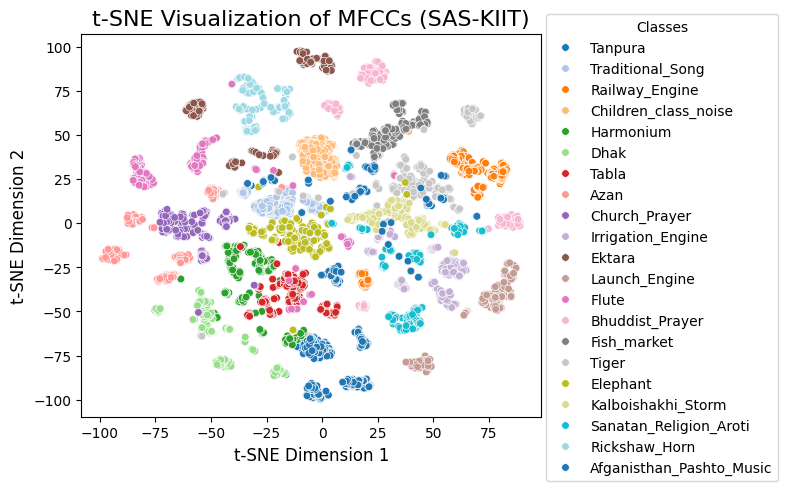}
        \caption{t-SNE: A Representation of Class Distributions in the Feature Space for SAS-KIIT Dataset.}
        \label{fig:1}
    \end{minipage}\hfill
\end{figure}

To better understand the dataset’s structure,  a t-SNE \cite{article} visualization (Fig. \ref{fig:1}) is provided that highlights class distinctions and overlaps, offering valuable insights into the dataset's distribution and aiding model interpretation before training.

\vspace{-1mm}

\vspace{-1mm}
\subsection{\textbf{UrbanSound8K Dataset}}
UrbanSound8K serves as a popular dataset in the field of environmental sound recognition, offering a wide range of urban noise samples. It includes various sound categories captured from real-world settings, making it suitable for evaluating classification models in complex acoustic environments.

\vspace{1mm}
\subsubsection{Dataset Overview}
The UrbanSound8K dataset consists of 1,302 labeled audio recordings, all of which were collected from freesound\footnote{www.freesound.org}, an open-source platform for audio sharing. Each recording is carefully annotated with start and end times. The dataset maintains the original audio properties, including codec, sampling rate, and bit depth, ensuring that the recordings retain their authenticity and quality without modifications. The dataset comprises 10 distinct sound classes, outlined as follows:

\vspace{-2.5mm}
\begin{table}[h]
    \caption{List of 10 Sound Classes in the UrbanSound8K Dataset}
    \centering
    \begin{tabular}{ll}
        \hline
        1. Air Conditioner & 6. Engine Sound \\
        2. Vehicle Horn & 7. Gun Shot \\
        3. Children Playing & 8. Jackhammer \\
        4. Dog Bark & 9. Siren \\
        5. Drilling & 10. Music \\
        \hline
    \end{tabular}
    \label{tab:urbansound_classes}
\end{table}

\subsubsection{Organization and Characteristics}
The UrbanSound8K dataset is organized into a 10-fold structure, facilitating standardized evaluation through cross-validation. Each sound recording is segmented into short clips of $\leq 4$ seconds, resulting in a total of 8,732 labeled segments. To ensure consistency in audio processing, all recordings have been resampled to a standard 44.1 kHz sampling rate. Additionally, a metadata file is provided, containing essential information such as file names, class labels, segment start and end times, and fold assignments.

\begin{figure}[H]
    \centering
    \begin{minipage}[]{0.48\textwidth} 
        \centering
        \includegraphics[width=\textwidth, height=6.3cm]{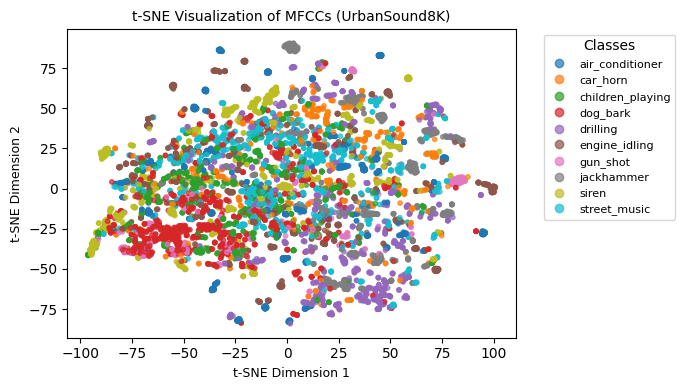}
        \caption{t-SNE: A Representation of Class Distributions in the Feature Space for UrbanSound8K Dataset.}
        \label{fig:2}
    \end{minipage}\hfill
\end{figure}

To further illustrate the dataset’s structure, a t-SNE (refer to Fig. \ref{fig:2}) visualization is provided, showcasing class distribution, relationships, and potential overlaps, which offer insights for model development and evaluation.
\vspace{-3mm}

\subsection{\textbf{Audio Mixing Process}}
\begin{figure}[H]
  \centering
  \includegraphics[width=10cm, height=4.8cm]{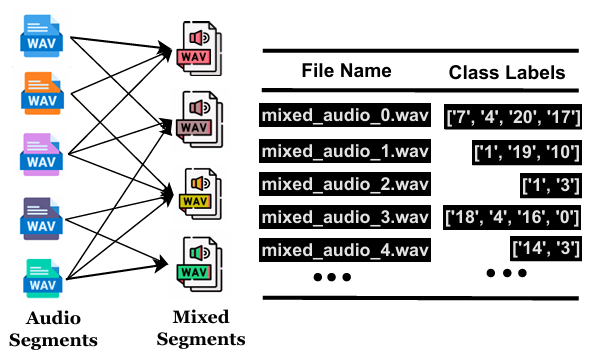}
  \caption{Audio Mixing Process and Label Interactions.}
  \label{fig:3}
\end{figure}

In this stage, mixed audio files (refer to Fig. \ref{fig:3}) are created by combining a fixed and variable number of segments. For fixed three and for variable each mixed file is constructed by merging between one and four individual audio segments. The total number of mixed audio samples generated is 8,000. This variability helps simulate real-world audio mixtures, making the dataset more robust. The mixed files are randomly distributed across 10 folders. Some mixed files can be found for reference\footnote{Available at: https://multilabelaudioclassification.netlify.app/}. A metadata file is maintained which is crucial for subsequent steps, as it facilitates the linking of the mixed files to their corresponding class labels and folder assignments.

\section{\textbf{Proposed Methodology}}\label{sec4}
\subsection{\textbf{Mel Spectrogram Generation}}
Each audio clip was sampled at 44.1 kHz and processed using 128 Mel filter banks with a maximum frequency of 8000 Hz. These parameters offer a good trade-off between frequency resolution and computational efficiency, allowing the model to capture important audio features while maintaining training speed. In total, 8000 Mel spectrograms were generated, corresponding to 8000 mixed audio samples. These spectrograms were stored as PNG images distributed across 10 folders, and a corresponding metadata file was maintained to track file properties and labels.

To convert each audio signal into a time-frequency representation, the Short-Time Fourier Transform (STFT) was used to convert each audio signal into a time-frequency representation, enabling the analysis of localized frequency changes over time. Mathematically, STFT is defined as:

\begin{equation}
\text{STFT}\{s[q]\}(p, \theta) = \sum_{q = -\infty}^{\infty} s[q] \cdot h[q - p] \cdot e^{-j \theta q}
\end{equation}

where \( s[q] \) represents the discrete-time audio signal, \( h[q - p] \) is the analysis window function centered at frame \( p \), and \( \theta \) denotes the angular frequency component. To generate the spectrogram \( S \), the magnitude of the STFT output is squared, as shown below:

\begin{equation}
S(p, \theta) = \left| \text{STFT}\{s[q]\}(p, \theta) \right|^2
\end{equation}

To align the frequency representation with human auditory perception, the spectrogram was processed using a Mel-scale filter bank, which maps the frequency axis onto a perceptually motivated scale, emphasizing regions more sensitive to the human ear. The result is a Mel spectrogram, which was then used as input to our CNN-based classification model.

\subsection{\textbf{Computation of MFCC Features}}
Mel-Frequency Cepstral Coefficients (MFCCs) were extracted from each audio clip sampled at 44.1 kHz. A total of 40 coefficients were computed per frame to represent the audio’s spectral characteristics. To maintain uniform input size, each MFCC matrix was either padded or truncated to 400 frames. Standardization was applied to normalize the features, improving consistency and training performance. These feature vectors were then paired with their corresponding multilabel annotations as defined in the metadata.

The MFCCs were derived by performing a Discrete Cosine Transform (DCT) on the log-compressed outputs of the Mel-scale filter bank as shown below:

\vspace{-2mm}
\begin{equation}
\text{MFCC}(k) = \sum_{u=1}^{U} \log(R(u)) \cdot \cos\left[\frac{\pi k}{U}(u - 0.5)\right]
\end{equation}

where \( R(u) \) represents the energy of the \( u \)-th Mel filter, and \( U \) is the total number of filters used.

\subsection{\textbf{Input Preparation}}
The Mel-spectrogram dataset was partitioned into three subsets following a 70-20-10 ratio for training, validation, and testing, respectively. The training subset facilitated model optimization, while the validation subset was utilized to monitor intermediate performance and mitigate overfitting risks. The test set served as an unbiased benchmark for final evaluation. Spectrogram images were resized to $128 \times 128$ pixels and transformed into tensors to ensure compatibility with the input requirements of the classification model.

\subsection{\textbf{Model Architecture}}

\begin{figure*}[ht]
    \centering
    \begin{minipage}[t]{0.95\textwidth}
        \centering
        \includegraphics[width=\textwidth, height=6.1cm]{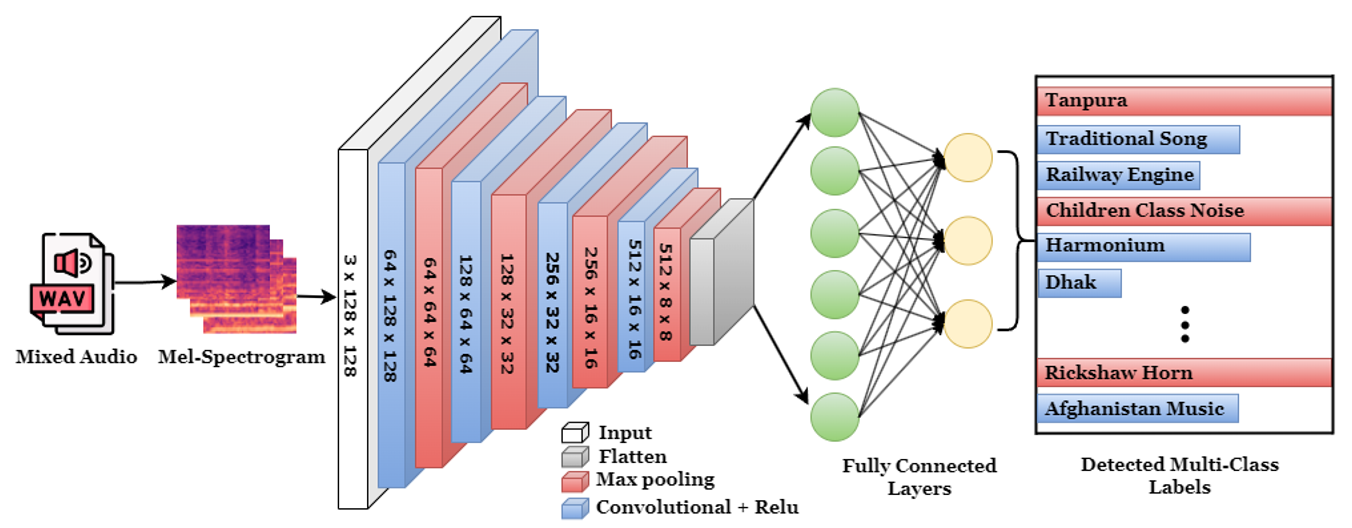} 
        \caption{Mixed audio is converted to Mel-spectrograms and fed into a CNN with ReLU and max pooling to extract features; dense layers output multilabel predictions, with dusty pink highlighting detected sound classes in an example recording.}
        \label{fig:4}
    \end{minipage}\hfill
\end{figure*}

A CNN-based framework was adopted to interpret spectrogram images for recognizing multiple co-occurring sound classes. The network integrates convolutional layers to extract relevant features, pooling layers to reduce spatial dimensions, and dense layers at the end for final class prediction, as illustrated in Figure~\ref{fig:4}. The architecture includes the following components:

\begin{itemize}
    \item \textbf{Convolutional Layers:}
    The architecture begins with 64 filters, followed by layers with 128, 256, and 512 filters, all using $3 \times $3 kernels and ReLU activation to extract progressively deeper spectrogram features.
    
    \item \textbf{Pooling Layers:} Max-pooling layers are applied after convolutional blocks, reducing spatial dimensions from $128 \times 128$ to $64 \times 64$ after the first pooling and to $32 \times 32$ after the last and so on, preserving key features while minimizing overfitting.

    \item \textbf{Fully Connected Layers:} The flattened feature maps are passed through a fully connected layer with 128 neurons (ReLU) and a final output layer with $C$ neurons (where $C$ is the number of classes), producing raw logits.

    \item \textbf{Activation Functions:} ReLU is used for intermediate layers, while the output layer produces logits, which are processed by a sigmoid function during loss computation.

    \item \textbf{BCEWithLogitsLoss:} Combines sigmoid activation and binary cross-entropy loss in one operation, eliminating the need for explicit sigmoid activation in the output layer. It ensures proper handling of multi-label classification tasks.The equations for the sigmoid function and the BCEWithLogitsLoss are provided below:
\end{itemize}

\begin{equation}
\sigma(\hat{p}_i) = \frac{1}{1 + e^{-\hat{p}_i}}
\end{equation}
The sigmoid outputs are utilized to compute the binary cross-entropy loss, as illustrated below:

\begin{equation}
\mathcal{L}_{\text{BCE}} = - \frac{1}{L} \sum_{j=1}^{L} \left[ p_j \cdot \log(\sigma(\tilde{p}_j)) + (1 - p_j) \cdot \log(1 - \sigma(\tilde{p}_j)) \right]
\end{equation}

In this equation, $\tilde{p}_j$ denotes the predicted logit for the $j$-th label, and $p_j$ is its corresponding ground truth. The sigmoid function $\sigma(\cdot)$ maps the logits to probabilities. The variable $L$ represents the total number of output labels. This formulation is ideal for multi-label classification tasks, where each label is treated independently during training.

The model is optimized to efficiently learn meaningful representations from spectrograms, ensuring robust classification performance.

\subsection{\textbf{Training Procedure}}
Model training was conducted over 100 epochs using a mini-batch size of 16. Optimization was handled by the Adam algorithm, initialized with a learning rate of 0.001. This setup balanced computational efficiency and stable gradient updates. The validation process assessed generalization performance, where spectrogram images were processed to generate raw logits. A threshold was applied to obtain binary labels for multi-label classification.

\section{\textbf{Result Analysis}}\label{sec5}
The proposed model is evaluated using both MFCC and Mel
Spectrogram features across two datasets. Each dataset was tested with different data splitting strategies to assess their impact on classification performance. To evaluate performance, precision(P), recall(R), and F1-score are used, ensuring a detailed assessment of class-wise classification. Additionally, accuracy was calculated based on thresholded predictions to measure performance.

\subsection{Results from fixed-mix audio dataset}
As mentioned earlier, the subset contains mixed audio samples created by combining three distinct sound sources. On SAS-KIIT and UrbanSound8K, the spectrogram-driven model produced accuracies of \textbf{95.42\%} and \textbf{86.36\%}, respectively. Table~\ref{tab:feature_comparison_fixed} provides a summary of the findings.

\begin{table}[h]
    \centering
    \caption{Performance comparison of MFCC and Spectrogram
    features on datasets with fixed no. mixed audio samples.}
    \renewcommand{\arraystretch}{1.45} 
    \setlength{\tabcolsep}{6.5pt} 
    \setlength{\arrayrulewidth}{1pt} 
    \begin{tabular}{|p{2.5cm} | p{2cm} | p{1cm} | p{1cm} | p{1cm} | p{1cm}|}
        \hline
        \cellcolor{blue!20} \textbf{Dataset} & 
        \cellcolor{yellow!20} \textbf{Feature} & 
        \cellcolor{red!20} \textbf{P} & 
        \cellcolor{teal!20} \textbf{R} & 
        \cellcolor{lime!20} \textbf{F1} & 
        \cellcolor{green!20} \textbf{Acc.} \\
        \hline
        {\centering SAS-KIIT} 
        & MFCC        & 0.83  & 0.73  & 0.76  & 93.91  \\
        \cline{2-6}
        & Spectrogram & 0.87  & 0.78  & 0.81  & \textbf{95.42} \\
        \hline
        {\centering UrbanSound8K} 
        & MFCC        & 0.71 & 0.61 & 0.62  & 84.16  \\
        \cline{2-6}
        & Spectrogram & 0.72  & 0.67  & 0.64  & \textbf{86.36} \\
        \hline
    \end{tabular}
    \label{tab:feature_comparison_fixed}
\end{table}

\subsection{Results from a Variable no. mix audio dataset}
In the variable mixed audio dataset, where one to four audio samples were combined to create a single mixed file is utilized. The results showed that the model, using spectrogram features, achieved an accuracy of \textbf{96.37\%} on the SAS-KIIT dataset and \textbf{85.26\%} on the UrbanSound8K dataset. As observed in the Table \ref{tab:feature_comparison_variable}, the model consistently performed well across various configurations of mixed samples, highlighting its capacity to process complex, real-world audio scenarios effectively. 

\begin{table}[h]
    \centering
    \caption{Performance comparison of MFCC and Spectrogram features on datasets with variable no. mixed audio samples.}
    \renewcommand{\arraystretch}{1.45} 
    \setlength{\tabcolsep}{6.5pt} 
    \setlength{\arrayrulewidth}{1pt} 
    \begin{tabular}{|p{2.5cm} | p{2cm} | p{1cm} | p{1cm} | p{1cm} | p{1cm}|}
        \hline
        \cellcolor{blue!20} \textbf{Dataset} & 
        \cellcolor{yellow!20} \textbf{Feature} & 
        \cellcolor{red!20} \textbf{P} & 
        \cellcolor{teal!20} \textbf{R} & 
        \cellcolor{lime!20} \textbf{F1} & 
        \cellcolor{green!20} \textbf{Acc.} \\
        \hline
        {\centering SAS-KIIT} 
        & MFCC        & 0.84  & 0.76  & 0.77  & 94.63  \\
        \cline{2-6}
        & Spectrogram & 0.89  & 0.83  & 0.84  & \textbf{96.37} \\
        \hline
        {\centering UrbanSound8K} 
        & MFCC        & 0.68  & 0.65  & 0.62  & 83.94 \\
        \cline{2-6}
        & Spectrogram & 0.69  & 0.63  & 0.61  & \textbf{85.26} \\
        \hline
    \end{tabular}
    \label{tab:feature_comparison_variable}
\end{table}

\begin{figure*}[ht]
    \centering
    \begin{minipage}[t]{0.95\textwidth}
        \centering
        \includegraphics[width=16.2cm, height=6.3cm]{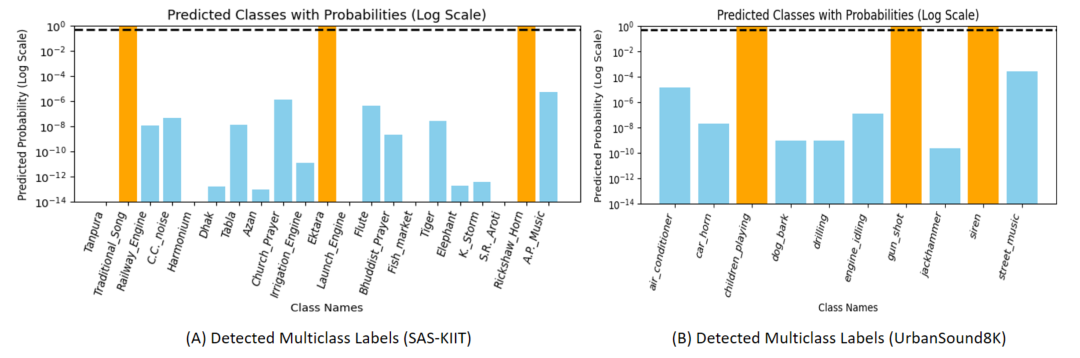} 
        \caption{Log-Scale Representation of Predicted Audio Classes: Orange bars indicate classes detected in the sample audio; Sky-blue bars show other predicted classes with lower probability scores in log scale; The dashed black line (- - -) represents the decision threshold, above which classes are considered present.}
        \label{fig:5}
    \end{minipage}\hfill
\end{figure*}

The t-SNE visualization (refer to Figure \ref{fig:2}) clearly shows that the UrbanSound8K dataset exhibits a more complex and overlapping class distribution compared to the SAS-KIIT dataset. This higher complexity makes classification more challenging, leading to a lower performance score on UrbanSound8K.
Figure \ref{fig:5} demonstrates how the model accurately detects and distinguishes multiple sound classes in a mixed audio input, supporting its practical effectiveness in real-world scenarios.

 \subsection{Comparative Analysis with Leading Techniques}
To evaluate the performance of the proposed model, comparisons were made with leading state-of-the-art (SOTA) approaches, both the FACE\cite{morsali2023face} and Pretrained Audio Neural Networks (PANNs) \cite{9229505} model were employed using spectrogram and mfcc
features extracted from the variable mixed audio dataset. PANNs are well-regarded for their robust performance in general-purpose audio classification tasks, having been trained on extensive and diverse datasets. FACE, a more recent and complex architecture, is specifically designed for environmental sound recognition. As presented in Table \ref{tab:feature_comparison_sota}, the proposed model surpasses FACE in terms of classification accuracy while maintaining a significantly simpler architecture. Compared to PANNs, it also delivers competitive performance with reduced computational requirements.

\begin{table}[h]
    \centering
    \caption{Performance comparison of the Proposed Model, FACE and PANNs on datasets using spectrogram features with variable number of mixed audio samples.}
    \renewcommand{\arraystretch}{1.45} 
    \setlength{\tabcolsep}{6.5pt} 
    \setlength{\arrayrulewidth}{1pt} 
    \begin{tabular}{|p{3cm} | p{3cm} | p{1cm} | p{1cm} | p{1cm} | p{1cm}|}
        \hline
        \cellcolor{blue!20} \textbf{Dataset} & 
        \cellcolor{yellow!20} \textbf{Model} & 
        \cellcolor{red!20} \textbf{P} & 
        \cellcolor{teal!20} \textbf{R} & 
        \cellcolor{lime!20} \textbf{F1} & 
        \cellcolor{green!20} \textbf{Acc.} \\
        \hline
        \multirow{3}{*}{\shortstack{SAS-KIIT}} 
        & FACE\cite{morsali2023face}  & 0.91  & 0.79  & 0.80  & 95.22  \\
        \cline{2-6}
        & PANNs\cite{9229505} & 0.85  & 0.76  & 0.78  & 92.51 \\
        \cline{2-6}
        & Proposed Model & 0.89  & 0.83  & 0.84  & \textbf{96.37} \\
        \hline
        \multirow{3}{*}{\shortstack{Urban-Sound8K}}
        & FACE\cite{morsali2023face}  & 0.71  & 0.57  & 0.59  & 84.54 \\
        \cline{2-6}
        & PANNs\cite{9229505} & 0.63  & 0.61  & 0.66  & 83.26 \\
        \cline{2-6}
        & Proposed Model & 0.69  & 0.63  & 0.61  & \textbf{85.26} \\
        \hline
    \end{tabular}
    \label{tab:feature_comparison_sota}
\end{table}

The results in Table \ref{tab:feature_comparison_sota} show that, despite FACE's complex design, our model outperforms it in accuracy while being faster and less complex. The model delivers higher classification accuracy while maintaining computational efficiency, making it well-suited for deployment in practical scenarios and environments with limited resources. This comparison highlights our model's high performance and practical advantages.

\section{\textbf{Conclusion}}\label{sec6}
The proposed convolutional neural network shows strong performance in multi-label classification of complex, mixed environmental sounds. By using spectrograms as input features, it effectively captures essential time-frequency characteristics, enabling accurate recognition even when multiple sources overlap within a single audio sample. The approach maintains high accuracy across culturally diverse and general-purpose datasets, demonstrating its versatility and reliability. Its simplicity and efficiency make it well-suited for urban monitoring, cultural soundscape preservation, and other real-world applications where environmental sound classification is crucial. Looking ahead, the model can be enhanced by incorporating advanced components such as attention mechanisms or temporal sequence modeling to better capture contextual dependencies. Its adaptability also offers potential for deployment on resource-constrained devices for real-time field analysis. Future work may further expand the framework with multi-modal data inputs to enrich performance in complex auditory scenes.

\bibliographystyle{IEEEtran}
\bibliography{example}
\end{document}